\documentclass[aps,prl,twocolumn,superscriptaddress,showpacs,preprintnumbers,amsmath,amssymb]{revtex4}
\usepackage{amssymb}
\usepackage{overpic}
\usepackage{graphicx} 
\usepackage{dcolumn}  
\usepackage{color}
\graphicspath{{ps}}

\begin{document}



\title{ \quad\\[1.0cm] Search for the $X(1812)$ in $B^{\pm} \rightarrow K^{\pm} \omega
\phi$}

\affiliation{Budker Institute of Nuclear Physics, Novosibirsk}
\affiliation{University of Cincinnati, Cincinnati, Ohio 45221}
\affiliation{T. Ko\'{s}ciuszko Cracow University of Technology,
Krakow} \affiliation{Department of Physics, Fu Jen Catholic
University, Taipei} \affiliation{Justus-Liebig-Universit\"at
Gie\ss{}en, Gie\ss{}en} \affiliation{The Graduate University for
Advanced Studies, Hayama}
\affiliation{Hanyang University, Seoul} \affiliation{University of
Hawaii, Honolulu, Hawaii 96822} \affiliation{High Energy Accelerator
Research Organization (KEK), Tsukuba} \affiliation{Hiroshima
Institute of Technology, Hiroshima}
\affiliation{Institute of High Energy Physics, Chinese Academy of
Sciences, Beijing} \affiliation{Institute of High Energy Physics,
Vienna} \affiliation{Institute of High Energy Physics, Protvino}
\affiliation{Institute for Theoretical and Experimental Physics,
Moscow} \affiliation{J. Stefan Institute, Ljubljana}
\affiliation{Kanagawa University, Yokohama} \affiliation{Korea
University, Seoul}
\affiliation{Kyungpook National University, Taegu}
\affiliation{\'Ecole Polytechnique F\'ed\'erale de Lausanne (EPFL),
Lausanne} \affiliation{Faculty of Mathematics and Physics,
University of Ljubljana, Ljubljana} \affiliation{University of
Maribor, Maribor} \affiliation{University of Melbourne, School of
Physics, Victoria 3010} \affiliation{Nagoya University, Nagoya}
\affiliation{Nara Women's University, Nara} \affiliation{National
Central University, Chung-li} \affiliation{National United
University, Miao Li} \affiliation{Department of Physics, National
Taiwan University, Taipei} \affiliation{H. Niewodniczanski Institute
of Nuclear Physics, Krakow} \affiliation{Nippon Dental University,
Niigata} \affiliation{Niigata University, Niigata}
\affiliation{University of Nova Gorica, Nova Gorica}
\affiliation{Novosibirsk State University, Novosibirsk}
\affiliation{Osaka City University, Osaka}
\affiliation{Panjab University, Chandigarh}
\affiliation{RIKEN BNL Research Center, Upton, New York 11973}
\affiliation{Saga University, Saga} \affiliation{University of
Science and Technology of China, Hefei} \affiliation{Seoul National
University, Seoul}
\affiliation{Sungkyunkwan University, Suwon} \affiliation{University
of Sydney, Sydney, New South Wales}
\affiliation{Toho University, Funabashi} \affiliation{Tohoku Gakuin
University, Tagajo}
\affiliation{Department of Physics, University of Tokyo, Tokyo}
\affiliation{Tokyo Metropolitan University, Tokyo}
\affiliation{Tokyo University of Agriculture and Technology, Tokyo}
\affiliation{IPNAS, Virginia Polytechnic Institute and State
University, Blacksburg, Virginia 24061} \affiliation{Yonsei
University, Seoul}
  \author{C.~Liu}\affiliation{University of Science and Technology of China, Hefei} 
  \author{Z.~P.~Zhang}\affiliation{University of Science and Technology of China, Hefei} 
  \author{I.~Adachi}\affiliation{High Energy Accelerator Research Organization (KEK), Tsukuba} 
  \author{H.~Aihara}\affiliation{Department of Physics, University of Tokyo, Tokyo} 
  \author{K.~Arinstein}\affiliation{Budker Institute of Nuclear Physics, Novosibirsk}\affiliation{Novosibirsk State University, Novosibirsk} 
  \author{T.~Aushev}\affiliation{\'Ecole Polytechnique F\'ed\'erale de Lausanne (EPFL), Lausanne}\affiliation{Institute for Theoretical and Experimental Physics, Moscow} 
  \author{A.~M.~Bakich}\affiliation{University of Sydney, Sydney, New South Wales} 
  \author{E.~Barberio}\affiliation{University of Melbourne, School of Physics, Victoria 3010} 
  \author{A.~Bay}\affiliation{\'Ecole Polytechnique F\'ed\'erale de Lausanne (EPFL), Lausanne} 
  \author{V.~Bhardwaj}\affiliation{Panjab University, Chandigarh} 
% ...
\collaboration{The Belle Collaboration}


\begin{abstract}
We report on a search for the $X(1812)$ state in the decay $B^{\pm}
\rightarrow K^{\pm} \omega \phi$ with a data sample of
$657\times10^6$ $B\overline{B}$ pairs collected with the Belle
detector at the KEKB $e^+e^-$ collider. No significant signal is
observed. An upper limit ${\cal B}(B^{\pm} \rightarrow K^{\pm}
X(1812),X(1812) \rightarrow \omega \phi)<3.2\times 10^{-7}$~($90\%$
C.L.) is determined. We also constrain the three-body decay
branching fraction to be ${\cal B}(B^{\pm} \rightarrow K^{\pm}
\omega \phi)$ $<$ 1.9 $\times 10^{-6}$~($90\%$ C.L.).

\end{abstract}


\maketitle

\tighten

{\renewcommand{\thefootnote}{\fnsymbol{footnote}}}
\setcounter{footnote}{0}

Using a sample of $5.8 \times 10^7~J/\psi$ events, the BES
collaboration observed a near-threshold enhancement in the $\omega
\phi$ invariant mass spectrum from the double OZI suppressed $J/\psi
\rightarrow \gamma \omega \phi$ decay with a statistical
significance of more than $10 \sigma$~\cite{x1812found}. When fitted
with a Breit-Wigner, this enhancement, called X(1812), has the
following mass, width, and product of branching fractions:
\begin{displaymath}
\begin{array}{c}
\label{xmass}M=(1812^{+19}_{-26}\pm 18) ~\textnormal{MeV/c$^{2}$} ,\\
 \Gamma=(105 \pm 20 \pm 28) ~\textnormal{MeV/c$^{2}$},\\
{\cal B}(J/\psi \rightarrow \gamma X,X \rightarrow \omega \phi)
=(2.60 \pm 0.27 \pm 0.65) \times 10^{-4}.
\end{array}
\end{displaymath}
Partial wave analysis favors a spin-parity assignment of
$J^{PC}=0^{++}$ for the $X(1812)$. In the related $\omega \psi$
mode, Belle has seen a dramatic threshold enhancement in $B^+
\rightarrow K^+ \omega \psi$, the Y(3940)~\cite{Abe:2004zs}, which
has now been confirmed by BaBar~\cite{Aubert:2007vj}.

 If the $X(1812)$ is a $q \overline q$ meson, the $X(1812) \rightarrow
\omega \phi$ branching fraction should be very small due to OZI
suppression and the limited available phase space, in contrast with
the BES observation. Suggestions have been made that the $X(1812)$
may be a tetraquark state (with structure $Q^2\overline{Q}{}^2$),
since some tetraquark states decay to vector-vector mesons
dominantly by ``falling apart'' and their masses are at the
threshold of two vector mesons~\cite{x4q}. Other works speculate
that it may be a hybrid~\cite{Chao:2006fq}, glueball
state~\cite{Bicudo:2006sd}, an effect due to intermediate meson
rescatterings~\cite{Zhao:2006dv} or a threshold cusp attracting a
resonance~\cite{Bugg:2008wu}. In this paper, we report our search
for this state in the decay $B^{\pm} \rightarrow K^{\pm} \omega
\phi$. On the other hand, this decay proceeds via a $b \rightarrow
s$ penguin with $s \overline{s}$ and $u \overline{u}$ popping. A
similar decay mode $B^+ \rightarrow K^+ \phi \phi$, which proceeds
via a $b \rightarrow s$ penguin diagram with double $s\overline{s}$
popping, is the only observed charmless $B \rightarrow VVP$ (two
vector mesons and one pseudoscalar meson) mode and has a rather
large branching fraction
[$(4.9^{+2.4}_{-2.2})\times10^{-6}$]~\cite{phiphik1,phiphik2}.
Therefore, even if the $X(1812)$ cannot be observed, measurement of
the $B^{\pm} \rightarrow K^{\pm} \omega \phi$ three-body decay is
also helpful for investigating decay mechanisms.

This analysis uses 605~fb$^{-1}$ of data containing $657 \times
10^6$ $B \overline{B}$ pairs. The data was collected with the Belle
detector~\cite{Belle} at the KEKB~\cite{KEKB} $e^{+} e^{-}$
asymmetric-energy (3.5~GeV on 8.0~GeV) collider operating at a
center-of-mass~(CM) energy of the $\Upsilon(4S)$ resonance.

The Belle detector is a large-solid-angle spectrometer~\cite{Belle}.
It consists of a silicon vertex detector~(SVD), a 50-layer central
drift chamber~(CDC), an array of aerogel threshold Cherenkov
counters~(ACC), time-of-flight scintillation counters~(TOF), and an
electromagnetic calorimeter comprised of CsI(Tl) crystals located
inside a superconducting solenoid that provides a 1.5 T magnetic
field. An iron flux return located outside the coil is instrumented
to detect $K^0_L$ mesons and to identify muons~(KLM).

\emph{B}-daughter candidates are reconstructed from the decays
$\omega \rightarrow \pi^+ \pi^- \pi^0$ , $\phi \rightarrow K^+ K^-$
and $\pi^0 \rightarrow \gamma \gamma$. Charged tracks are identified
as pions or kaons by combining information from the CDC, ACC and TOF
systems. We reduce the number of poor quality tracks by requiring
that $\mid$$dr$$\mid$ $<0.3$~cm and $\mid$$dz$$\mid$ $<1.5$~cm,
where $\mid$$dr$$\mid$ and $\mid$$dz$$\mid$ are the distances of
closest approach of a track to the interaction point in the
transverse plane and \emph{z} direction (opposite to the direction
of the positron beam), respectively. In addition, tracks matched
with clusters in the ECL that are consistent with an electron
hypothesis are rejected. We use a kaon identification likelihood
ratio $R_{K,\pi}$ = $L_K$/($L_K$+$L_{\pi}$) to discriminate \emph{K}
and $\pi$ candidates. The requirements $R_{K,\pi}$$>$0.4 for a kaon
and $R_{K,\pi}$$<$0.6 for a $\pi$ are used. The efficiency to
identify a kaon(pion) is 94\%, while the probability that a
pion(kaon) is misidentified as a kaon(pion) is about 10\%. Candidate
$\pi^0$ mesons are reconstructed from pairs of photons, where the
energy of each photon in the laboratory frame is required to be
greater than 50 MeV. We select $\pi^0$ mesons with an invariant mass
in the range 0.1193~GeV/$c^2$ $<$ $M(\gamma \gamma)$ $<$
0.1477~GeV/$c^2$ and a momentum in the laboratory frame
$p^{\textnormal{lab}}_{\pi^0}$ $>$ 0.38~GeV/$c$.

Particles satisfying the above selection criteria are then used to
reconstruct $\omega$ and $\phi$ mesons.
We select candidates in the invariant mass windows 0.75~GeV/$c^2 <
M_{\pi^+ \pi^- \pi^0} <$ 0.81~GeV/$c^2$ and 1.00~GeV/$c^2 < M_{K^+
K^-} <$ 1.04~GeV/$c^2$. A vertex fit for the $\phi$ and $\omega$
candidates is also performed. In addition, we require three kaons in
the final state, one directly from the \emph{B}-meson decay and the
other two from the $\phi$ decay. To distinguish the two kinds of
kaons and reduce multiple candidates, we require kaons from the
$\phi$ to have momenta $p_{K^{\pm}}$ $<$ 1.5~GeV/$c$ in the CM
frame.

Candidate $B^{\pm} \rightarrow K^{\pm} \omega \phi$ decays are
identified by using the energy difference $(\Delta E)$ and the
beam-energy-constrained mass ($M_{\textnormal{bc}}$). These are
defined as $\Delta E \equiv E_B -E_{\textnormal{beam}}$ and
$M_{\textnormal{bc}} \equiv \sqrt{E^2_{\textnormal{beam}}-p^2_B}$,
where $E_{\textnormal{beam}}$ denotes the beam energy, $E_B$ and
$p_B$ denote the reconstructed energy and momentum of the candidate
\emph{B}-meson, all evaluated in the $e^+e^-$ CM frame. We select
events satisfying $\mid$$\Delta E$$\mid$ $<$ 0.2~GeV and
5.20~GeV/$c^2$ $<$ $M_{\textnormal{bc}}$ $<$ 5.29~GeV/$c^2$, and
define signal regions $-0.15$~GeV $<$ $\Delta E$ $<$ 0.05~GeV and
5.27~GeV/$c^2$ $<$ $M_{\textnormal{bc}}$ $<$ 5.29~GeV/$c^2$.

The dominant source of background arises from random combinations of
particles in continuum $e^+e^- \rightarrow q \overline{q}$ events
(\emph{q=u,d,s,c}). To discriminate spherical $B \overline{B}$
events from jet-like $q \overline{q}$ events, we use event-shape
variables: specifically, 16 modified Fox-Wolfram moments~\cite{sfw}
combined into a Fisher discriminant,
$\mathcal{F}$~\cite{foxwolfram}. Additional discrimination is
provided by $\theta_{B}$, the polar angle in the CM frame between
the \emph{B} direction and \emph{z} direction. Correctly
reconstructed \emph{B}-mesons follow a ($1-\cos^2\theta _B$)
distribution, while fake candidates from continuum tend to be
uniform in cos$\theta_B$.

Further continuum background suppression is achieved using
\emph{b}-flavor tagging information. The Belle flavor tagging
algorithm~\cite{ftag} yields the flavor of the tagged meson,
\emph{q}(=$\pm 1$), and a flavor-tagging quality factor, \emph{r}.
The latter ranges from zero for no flavor discrimination to one for
unambiguous flavor assignment. For signal events, \emph{q} is
usually consistent with the flavor opposite to that of the signal
\emph{B}, while it is random for continuum events. Thus, the
quantity $qrF_{B}$ is used to separate signal and continuum events,
where $F_B$ is the charge of the signal \emph{B}: $F_B$ =$+1~(-1)$
for $B^+~(B^-)$

We use a Monte Carlo (MC) sample~\cite{evtgen} to form $\mathcal{F}$
and to obtain the cos$\theta_B$ and $qrF_B$ distributions.
Probability density functions (PDFs) are derived from $\mathcal{F}$
and the cos$\theta_B$ distributions and are multiplied to form
signal ($L_s$) and continuum background $(L_{q \overline q})$
likelihood functions, which are further combined to form a
likelihood ratio $R_s=L_s/(L_s+L_{q \overline q})$. We divide events
into six $qrF_B$ bins and determine the optimum $R_s$ selection
criteria for each bin by maximizing $N_s/\sqrt{N_s+N_b}$, where
$N_s$ is the number of signal MC events in the signal region, and
$N_b$ is the number of background events estimated to be in the
signal region by assuming ${\cal B}(B^{\pm} \rightarrow K^{\pm}
\omega \phi)=1.0 \times 10^{-5}$. This optimization preserves 57.9\%
of the signal while rejecting 98.6\% of the continuum background.

Applying all of the above criteria, the fraction of events having
multiple candidates is 21\%. To select the $B$-meson candidate, we
add the $\chi^2$ of the $\omega$ meson vertex fit, and the $\chi^2$
of a $\pi^0 \rightarrow \gamma \gamma$ fit constrained to the
PDG~\cite{pdg} value of $\pi^0$ mass: the candidate with the
smallest value is chosen. If multiple \emph{B} candidates still
remain, we use the $\chi^2$ of the $\phi$ meson vertex fit and
$\chi^2$ of the \emph{B}-meson vertex fit to choose the best one.

\begin{figure}[htb]
\includegraphics[width=0.45\textwidth]{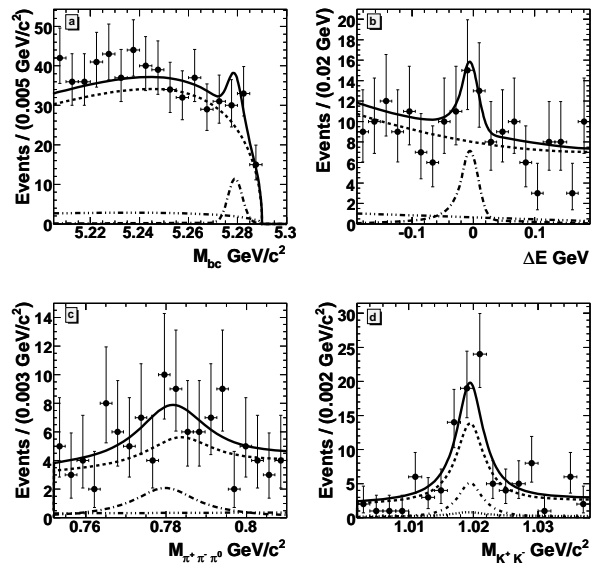}
\caption{ Projection of the data (points with error bars) and fit
results onto (a) $M_\textnormal{bc}$, (b) $\Delta E$, (c)
$M_{\pi^+\pi^-\pi^0}$, (d) $M_{K^+K^-}$ with the other variables
satisfying $M_\textnormal{bc} \in (5.27,5.29)~\textnormal{GeV}/c^2$,
$\Delta E \in (-0.15,0.05)~\textnormal{GeV}$, $M_{\pi^+ \pi^- \pi^0}
\in (-0.75,0.81)~\textnormal{GeV}/c^2$, $M_{K^+ K^-} \in (1.00,1.04)
~\textnormal{GeV}/c^2$ : signal (dot-dashed), $q \overline q$
(dashed), $B \overline B$ (dot-dot-dot-dashed) and total (solid).}
\label{result4d}
\end{figure}

In addition to the dominant continuum background, charmed \emph{B}
decay ($b \rightarrow c$) and charmless \emph{B} decay ($b
\rightarrow u, d, s$) backgrounds are studied using dedicated MC
samples that are respectively about two and 25 times the size of the
data sample. Charmless \emph{B} decay background is found to be
small and is neglected. The following charmed $B$ decay channels are
studied using dedicated Monte Carlo samples: $B^{\pm}\rightarrow
D_{s}^{\mp} \pi^0 \pi^{\pm}K^{\pm}$, $D_{s}^{\pm}\rightarrow
\pi^{\pm}\phi$ ; $B^{\pm}\rightarrow D_{s}^{\mp} \pi^{\pm}K^{\pm}$,
$D_{s}^{\pm}\rightarrow \pi^0 \pi^{\pm}\phi$ and $B^{\pm}\rightarrow
\overline{D}{}^0/D^0 K^{\pm}$, $D^0/\overline{D}{}^0\rightarrow
\pi^+ \pi^- \pi^0 K^+ K^-$. To measure the three-body  $B^{\pm}
\rightarrow K^{\pm} \omega \phi$ branching fraction, we require
$M_{K^+ K^-\pi^+ \pi^- \pi^0}$ $>$ 2.2~GeV/$c^2$ to exclude $D^0$
background, and require $\mid$$M_{\pi K^+
K^-}-m_{D_s}$$\mid$$>$0.15~GeV/$c^2$ as well as $\mid$$M_{\pi^0 \pi
K^+ K^-}-m_{D_s}$$\mid$$>$0.15~GeV/$c^2$, where $m_{D_s}$ is the
nominal $D_s$ mass~\cite{pdg}, to veto the $D_s$ background.

We obtain the signal yield using a four-dimensional extended
unbinned maximum likelihood (ML) fit to $\Delta E$,
$M_{\textnormal{bc}}$, $M_{\pi\pi\pi}$ and $M_{KK}$. The likelihood
function consists of the following components: signal decays,
continuum background ($q \overline q$), and charmed B-decay
background ($b \rightarrow c$). For all components, no sizable
correlations are found among the fitting quantities. The PDF for
event \emph{i} and component \emph{j} is defined as
\begin{eqnarray}
\mathcal{P}^i_j= \mathcal{P}_j(\Delta E^i)\times
\mathcal{P}_j(M^i_{\textnormal{bc}}) \times \mathcal{P}_j(M^i_{KK})
\times \mathcal{P}_j(M^i_{\pi \pi \pi}).
\end{eqnarray}

The signal $M_{\textnormal{bc}}$ is parameterized by the sum of a
single Gaussian and an ARGUS function~\cite{argus}, $\Delta E$ by a
Gaussian and a Crystal Ball function~\cite{crytalball}, and
$M_{K^+K^-}$, $M_{\pi^+ \pi^- \pi^0}$ by Breit-Wigner functions. For
continuum background, $M_{\textnormal{bc}}$ is  parameterized by an
ARGUS function, $\Delta E$ by a second-order Chebyshev polynomial,
and $M_{K^+K^-}$, $M_{\pi^+ \pi^- \pi^0}$ by the sum of Breit-Wigner
functions and first-order Chebyshev polynomials. $B \overline B$
background modelling is similar, but with a first-order Chebyshev
for $\Delta E$ and for $M_{\pi^+\pi^-\pi^0}$. All function
parameters are determined from MC simulation.

In our final fit to the data, the signal and $q \overline q$ yields
are allowed to vary; the fraction of $b \rightarrow c$ events is
very small and thus the yield is fixed in the fit according to MC.

The likelihood function to be maximized is given by
\begin{equation}
\label{yield} \mathcal{L}= \frac{e^{-(\sum_{j}
Y_j)}}{\emph{N}!}\prod_{i=1}^N{\sum_j{(Y_j\mathcal{P}_j^i)}}
\end{equation}
where $Y_j$ is the yield of events for component j and N is the
total number of events in the sample.

Figure~\ref{result4d} shows the fit results. Peaking behavior
observed in $\Delta E$, $M_{\textnormal{bc}}$, $M_{\pi \pi \pi }$
and $M_{KK}$ is consistent with that from MC expectations. The
branching fraction is evaluated using the following quantities: the
signal yield $Y_{\omega \phi K} = 22.1^{+8.3}_{-7.2}$ with
reconstruction efficiency $\varepsilon = 7.04\times 10^{-2}$; the
combined daughter branching fraction $\cal{B}$$_{\textnormal{d}}$
$=0.439$~\cite{pdg}; a correction of 0.946 to the efficiency of
$K/\pi$ identification requirements, which takes into account small
differences between MC and data; and a total of $657 \times 10^6$
produced $B \overline B$ pairs, where equal fractions of $B^+B^-$
and $B^0\overline{B}{}^0$ are assumed.

\begin{table}[htb]
\caption{ Systematic errors for ${\cal B}(B^{\pm} \rightarrow
K^{\pm} \omega \phi)$. In cases where the error on ${\cal B}(B^{\pm}
\rightarrow K^{\pm} X(1812),X(1812) \rightarrow \omega \phi)$ is
different, it is shown separately in parentheses. }\label{sys}
\begin{tabular}
{cccc} \hline \hline
\multicolumn{2}{c}{Type} & \multicolumn{2}{c}{Fractional error (\%)} \\
\multicolumn{2}{c}{}& $+\sigma$ & $-\sigma$ \\
\hline
\multicolumn{2}{l}{Tracking} & 6.00  & 6.00  \\
\multicolumn{2}{l}{$K/\pi$ ID}  & 2.90 & 2.90 \\
\multicolumn{2}{l}{$\pi^0$ Reconstruction} & 4.00 & 4.00 \\
\multicolumn{2}{l}{Daughter $\cal{B}$}  & 1.45 & 1.45\\
\multicolumn{2}{l}{Signal/Background Modeling} & 5.25(31.1) & 2.50(22.7)\\
\multicolumn{2}{l}{$B \overline B$ Background Yield} & 1.75(2.10) & 0.89(1.60)\\
\multicolumn{2}{l}{MC Statistics} & 0.56(1.06) & 0.56(1.06)\\
\multicolumn{2}{l}{Continuum Suppression} & 7.21(12.0) & 7.26(12.4)\\
\multicolumn{2}{l}{$N_{B \overline B}$} & 1.36(1.36) & 1.36(1.36)\\
\multicolumn{2}{l}{Total} & \bf{12.1(34.4)} & \bf{11.2(27.1)}\\
\hline \hline
\end{tabular}
\end{table}

The sources of systematic error are listed in Table ~\ref{sys}. The
quoted 6\% track reconstruction efficiency is from the consideration
that there are five tracks in a selected event and for each track
the efficiency error is 1.2\%. The errors due to continuum
suppression requirements are obtained by varying these cuts while
the errors on the PDF shapes are obtained by varying all fixed
parameters by $\pm 1 \sigma$. Toy MC tests and GEANT-based Detector
Simulation~(GSIM) tests are performed, we find that the fit bias can
be neglected. To estimate the error due to the $b \rightarrow c$
contribution, we vary the normalizations by $\pm 50 \%$.

Our final result for the three-body branching fraction based on the
605~fb$^{-1}$ data sample is
\begin{displaymath}
\label{brresult} {\cal B}(B^{\pm} \rightarrow K^{\pm} \omega
\phi)=(1.15 {+0.43 \atop -0.38} {+0.14 \atop -0.13} )\times 10^{-6},
\end{displaymath}
where the first error quoted is statistical and the second
systematic. We obtain the 90\% confidence level upper limit ${\cal
B}(B^{\pm} \rightarrow K^{\pm} \omega \phi)< 1.9 \times 10^{-6}$ by
a frequentist method using ensembles of pseudo-experiments. For a
given signal yield, 10000 sets of signal and background events are
generated according to the PDFs, and fits are performed. The
confidence level is obtained from the fraction of samples that give
a fit yield larger than that of data~(22.1). We take into account
systematic errors by varying the fit yield by the total systematic
errors described in Table~\ref{sys}. The significance of the signal,
estimated using this method, is $2.8 \sigma$.
\begin{figure}[htb]
\begin{overpic}[scale=.35]{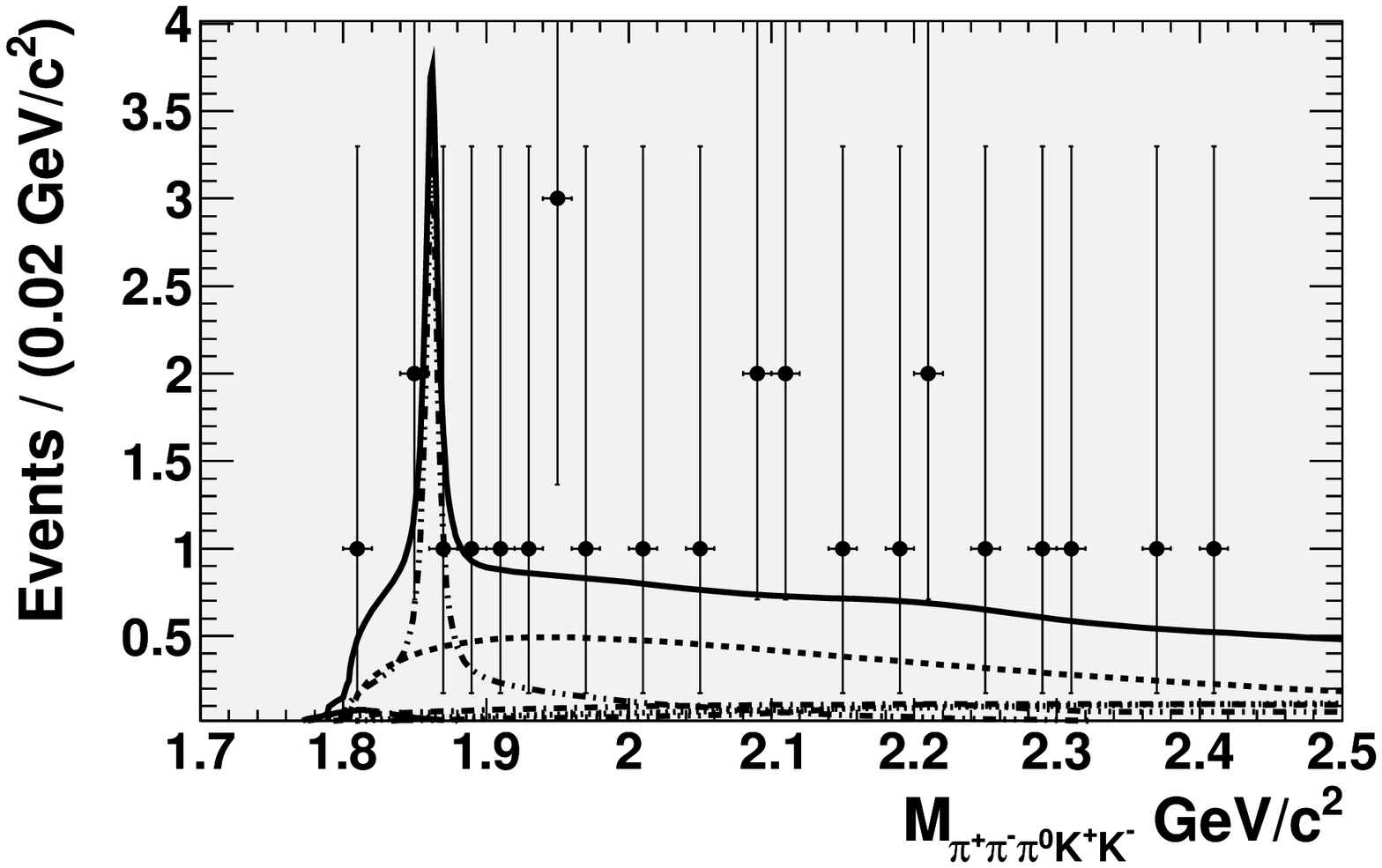}
\put(52,35){
  \includegraphics[scale=0.15]{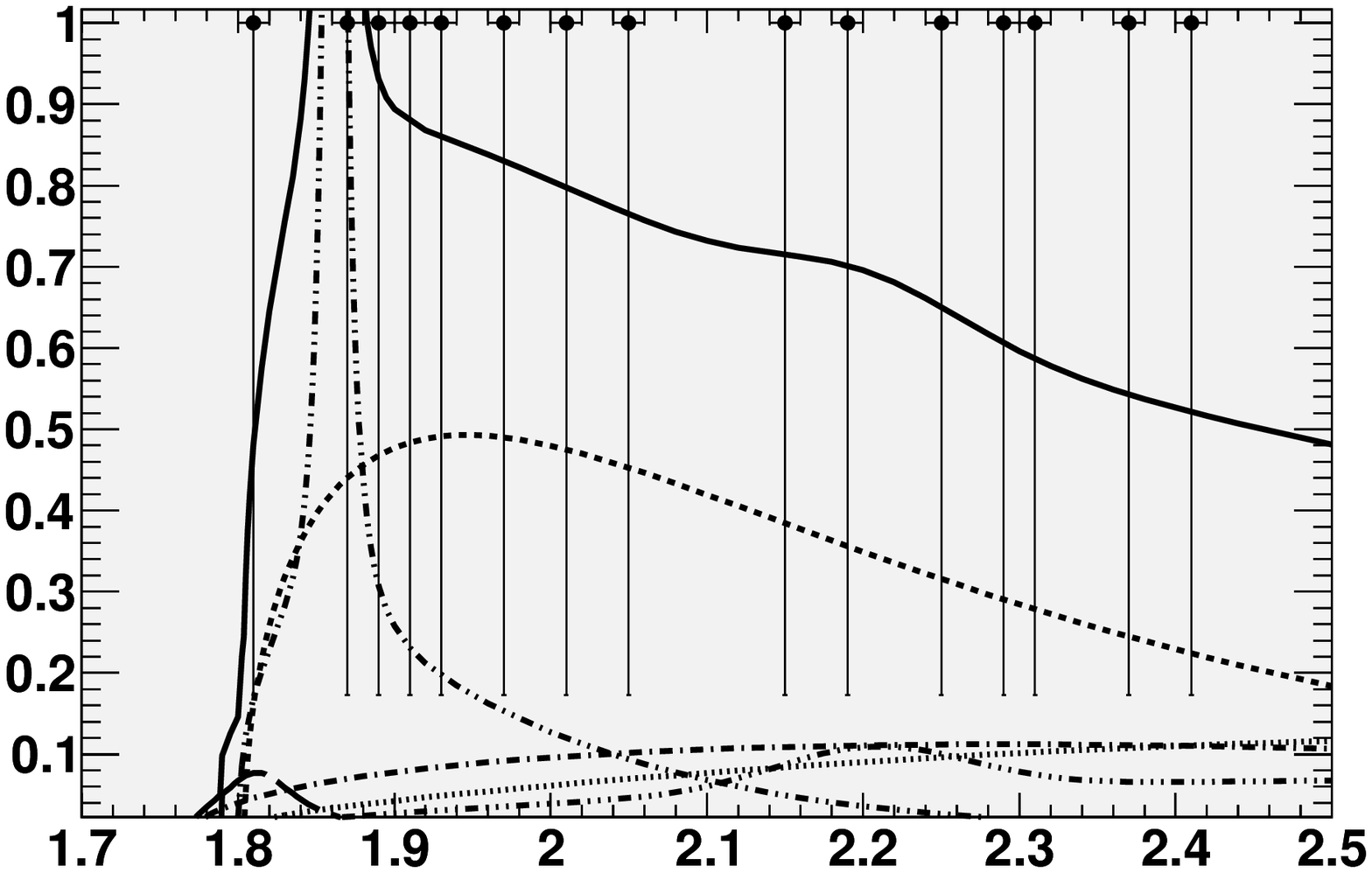}
  }
\end{overpic}
\caption{Mass spectrum in the $\omega \phi$ fit with the following
components: $B^+ \rightarrow K \omega \phi$ three-body (dotted), $B
\overline B$ (dot-dashed), $q \overline q$ (dashed),
$D^0$(dot-dot-dashed), $D_s$(dot-dot-dot-dashed), $B^{\pm}
\rightarrow K^{\pm} X(1812)$(long-dashed), and total(solid). The
spectrum is also shown in the inset with an expanded vertical scale}
\label{xfitresult}
\end{figure}

We next study the $\omega \phi$ mass spectrum. Because the
aforementioned $M_{K^+ K^-\pi^+ \pi^- \pi^0} $, $\mid$$M_{\pi K^+
K^-}-m_{D_s}$$\mid$ and $\mid$$M_{\pi^0 \pi K^+ K^-}-m_{D_s}$$\mid$
mass cuts influence the shape of the $\omega \phi$ invariant mass
spectrum, we did not use them and fit the $D^0$ and $D_{s}$
backgrounds simultaneously.

We produced 0.6 million (2.0 million) $D^0$ ($D_{s}$) background MC
events for the decay $B^{\pm} \rightarrow K^{\pm} X(1812),X(1812)
\rightarrow \omega \phi$. The $X(1812)$ mass and width are taken
from the BES measurement and its PDF is modeled by an rARGUS
(reversed ARGUS~\cite{argus},
$F_\textnormal{rARGUS}(x)=F_\textnormal{ARGUS}(2t-x)$, where $t$ is
the threshold) function plus a Breit-Wigner with a threshold. The
three-body decay PDF is an rARGUS function, the $D^0$ background PDF
is the sum of an rARGUS function and a Breit-Wigner, the $D_s$
background is the sum of an rARGUS function and a Gaussian, while
the $q \overline q$, $B \overline B$ backgrounds are also modeled by
rARGUS functions. We obtained all the parameters from MC samples. In
our final fit to the data, we fixed the yield of $D^0$ and $D_s$
backgrounds according to the PDG branching fractions~\cite{pdg}, and
fixed the yield of $B \overline B$ background according to MC
simulation.

The final result is shown in Fig.~\ref{xfitresult}. No significant
signal is observed; the yield of the $X(1812)$ is
$0.2_{-1.5}^{+2.4}$ events. The systematic errors are also listed in
Table.~\ref{sys}, where those in parentheses are for the items that
differ from those in the three-body decay analysis. We also include
the errors from the fraction of $D^{0}$, $D_{s}$ background and the
$X(1812)$ width into signal/background modeling. Using the
pseudo-experiment method described above and taking the systematic
errors into account, we find a limit on the product branching
fraction of ${\cal B}(B^{\pm} \rightarrow K^{\pm} X(1812),X(1812)
\rightarrow \omega \phi)<3.2\times 10^{-7}$~($90\%$ C.L.)

In summary, using a data sample of 605~fb$^{-1}$ collected with the
Belle detector, we present a search for the $X(1812)$ meson in the
decay $B^{\pm} \rightarrow K^{\pm} \omega \phi$. No significant
signal is observed. An upper limit for the product ${\cal B}(B^{\pm}
\rightarrow K^{\pm} X(1812),X(1812) \rightarrow \omega
\phi)<3.2\times 10^{-7}$~($90\%$ C.L.) is determined. We also
measure the three-body $B^{\pm} \rightarrow K^{\pm} \omega \phi$
decay branching fraction ${\cal B}(B^{\pm} \rightarrow K^{\pm}
\omega \phi)=[1.15 {+0.43 \atop -0.38} {+0.14 \atop -0.13}
(<1.9)]\times 10^{-6}$, where the upper limit is at the $90\%$
confidence level.

We thank the KEKB group for the excellent operation of the
accelerator, the KEK cryogenics group for the efficient operation of
the solenoid, and the KEK computer group and the National Institute
of Informatics for valuable computing and SINET3 network support.
We acknowledge support from the Ministry of Education, Culture,
Sports, Science, and Technology (MEXT) of Japan, the Japan Society
for the Promotion of Science (JSPS), and the Tau-Lepton Physics
Research Center of Nagoya University; the Australian Research
Council and the Australian Department of Industry, Innovation,
Science and Research; the National Natural Science Foundation of
China under contract No.~10575109, 10775142, 10875115 and 10825524;
the Department of Science and Technology of India; the BK21 program
of the Ministry of Education of Korea, the CHEP src program and
Basic Research program (grant No. R01-2008-000-10477-0) of the Korea
Science and Engineering Foundation; the Polish Ministry of Science
and Higher Education; the Ministry of Education and Science of the
Russian Federation and the Russian Federal Agency for Atomic Energy;
the Slovenian Research Agency;  the Swiss National Science
Foundation; the National Science Council and the Ministry of
Education of Taiwan; and the U.S.\ Department of Energy. This work
is supported by a Grant-in-Aid from MEXT for Science Research in a
Priority Area ("New Development of Flavor Physics"), and from JSPS
for Creative Scientific Research ("Evolution of Tau-lepton
Physics").

\end{document}